\documentclass[final,3p,times]{elsarticle}
\usepackage{url}
\usepackage{lineno,hyperref,xcolor}
\usepackage{graphicx}
\usepackage{caption}
\usepackage{booktabs}
\usepackage{multirow}
\usepackage{enumitem}
\biboptions{numbers,sort&compress}
\modulolinenumbers[5]
\hypersetup{colorlinks,% Reference color setup	
	linkcolor=Cerulean,%
citecolor=Cerulean}

\begin{document}

\begin{frontmatter}
	
	\title{Predictability of diffusion-based recommender systems}%\tnoteref{mytitlenote}}
	
	\author[mymainaddress]{Peng Zhang}
	
	\author[mymainaddress]{Leyang Xue}
	
	\author[mysecondaryaddress]{An Zeng\corref{mycorrespondingauthor}}
	\cortext[mycorrespondingauthor]{Correspondence and requests for materials should be addressed to A.Z}
	\ead{anzeng@bnu.edu.cn}

	\address[mymainaddress]{School of Science, Beijing University of Posts and Telecommunications, Beijing 100876, P.R. China.}
	\address[mysecondaryaddress]{School of Systems Science, Beijing Normal University, Beijing 100875, P.R. China.}
	
	\begin{abstract}
The recommendation methods based on network diffusion have been shown to perform well in both recommendation accuracy and diversity. Nowdays, numerous extensions have been made to further improve the performance of such methods. However, to what extent can items be predicted by diffusion-based algorithms still lack of understanding. Here, we mainly propose a  method to quantify the predictability of diffusion-based algorithms. Accordingly, we conduct experiments on Movielens and Netflix datasets. The results show that the higher recommendation accuracy based on diffusion algorithms can still be achieved by optimizing the way of resource allocation on a density network. On a sparse network, the possibility of improving accuracy is relatively low due to the fact that the current accuracy of diffusion-based methods is very close its predictability. In this case, we find that the predictability can be improved significantly by multi-steps diffusion, especially for users with less historical information. In contrast to common belief, there are plausible circumstances where the higher predictability of diffusion-based methods do not correspond to those users with more historical recording. Thus, we proposed the diffusion coverage and item average degree to explain this phenomenon. In addition, we demonstrate the recommendation accuracy in real online system is overestimated by random partition used in the literature, suggesting the recommendation in real online system may be a harder task.
	\end{abstract}

	\begin{keyword}
		\texttt{Predictability, Diffusion-based methods, Recommender system}
		%\MSC[2010] 00-01\sep  99-00
	\end{keyword}
	
\end{frontmatter}

%\linenumbers

\section{Introduction}

Due to the ongoing rapid expansion of internet, abundant information is available on the Iternet. This situation make it more difficult for users to select the items that look similar to what they are looking for. Therefore, it is significant to filter unrelated information and provide personalized recommendations to meet the manifold needs of the people. Nowdays, recommender system \cite{1,37} has become the essential tool for online life. Many e-commerce websites rely on the recommender system to suggest consumers new possibly relevant items, e.g. Amazon, Netflix, YouTube, Alibaba etc \cite{38,2}. 

Recommender systems are widely studied and is a mature research field. Various recommendation algorithms have been proposed to solve the information overload problem, including context-based analysis \cite{3,4}, collaborative filtering \cite{5,6}, latent semantic models \cite{7}, deep neural network\cite{8} and so on. In recent years, recommender system has attracted a large number of scholars from mathmatics and physics, and they have introduced some new ideas and proposed many widely accepted recommendation algorithms\cite{9,10,11,33}. Diffusion-based recommendation methods are such a class of algorithms that have received considerable attention in literature. For instance, mass diffusion (MD)\cite{10}, heat conduction(HC)\cite{9} and hybrid methods of both (Hyb)\cite{11} are personalized recommendation algorithms based on diffusion processes on bipartite networks\cite{12}. Besides, there are also plenty of extension methods which are designed to improve the recommendation performance, such as indirect-link-weakened mass diffusion method (IMD)\cite{13} ,non-equilibrium mass diffusion algorithm (NMD)\cite{14} and so on\cite{15,16,17,18,19,20,21,34}.

In previous studies, many diffusion-based methods have been proposed to accurately predict user's preferences. However, the resource diffusion process on bipartite network still raises fundamental questions: the extent to which the resource allocation process can improve the accuracy of recommendations is unknown. Similar problems are called predictability of link or algorithms by many scholars\cite{24,25}. Current studies with respect to predict ability mainly include: The predictability of nine link prediction algorithms based on common neighbor similarity in theory is analyzed by Xu et al\cite{24}. The predictability of network is treated as an inherent property of the network itself and can be reflected by structural consistency, which have been proposed by L\"u et al\cite{25}. Prior research focus on the predictability of network as well as specific link prediction algorithm. Here, we mainly study the predictability of diffusion-based algorithms, which (\romannumeral1) provides us with a comparison criterion to guide whether we should propose some new algorithms, (\romannumeral2) help us understand the limitations of diffusion-based algorithms, and (\romannumeral3) extract the skeleton structure of network to reduce computational complexity\cite{36}.

The intended contribution of this work is (\romannumeral1) propose to a method to quantify the predictability of diffusion-based recommendation algorithms; (\romannumeral2) show that existing diffusion-based algorithms have great potential to improve recommendation accuracy on a dense network, however it is ineffective to extend the existing diffusion-based algorithms to improve accuracy on a sparse network; (\romannumeral3) find that predictability of diffusion-based algorithms for users with small and big degree perform very poor and can be enhanced significantly by multi-step resource diffusion; (\romannumeral4) reveal the recommendation accuracy in real online is overestimated by random dataset partition. The paper is organized as follows. In \autoref{section2}, we will describe the resource diffusion process on bipartite network and introduce a method to quantify the predictability of diffusion-based algorithms. In \autoref{section3}, we will introduce the datasets, dataset partition, and evaluation metrics. In \autoref{section4}, we will present the results. In \autoref{section5}, we will conclude this work with a brief outlook of the future work.

\section{Methods}
\label{section2}

Online commercial system are naturally modeled as bipartite network where users and items are represented as nodes and an edge implies that an user has purchased an item. The bipartite network can be represented by a $M \times N$ matrix $A$ ($M$ users and $N$ items) where the element $A_{i\alpha} = 1$ if an user $i$ has selected an item $\alpha$ and $A_{i\alpha} = 0$ otherwise. Personalized recommendations based on diffusion methods for a target user are obtained by the following three steps\cite{1}:
\begin{itemize}
\item The initial resource of item is acquaired by the target user's historical preferences on a given bipartite network.
\item Each item distributes initial resource to all users who have collected it according to a resource allocation way.
\item Each user reallocate the resources to all items collected by the user according to a resources allocation way.  		
\end{itemize}

When the resource goes from one type of nodes to the other type of nodes, it is called a step of diffusion. Through the above three steps, the recommendation can be made. In fact, recommendation can also be based on diffusion with more than three steps. For instance, after five steps of diffusion, the resource reaches again the item side, with the ranking of which the recommendation list can be generated. 

Predictability is usually defined as the possible maximum accuracy of a recommendation algorithm\cite{35,25}. Intuitively, the predictability of recommendation algorithms should be 1 under such definition because those unselected items are always distinguishable. However, diffusion-based algorithms have fundamental limits on recommendation of items. For instance, some items cannot receive the resource from the target users under a given diffusion steps, which leads to these items are unpredictable. Therefore, the amount of items covered by resource is determined by the number of diffusion steps or intrinsic network structure. Inspired by this idea, the upper bound of recommendation accuracy of diffusion-based methods can be obtained at a given number of diffusion steps by a method called Ideal. The Ideal is described as follows: First, we suppose that there is an ideal resource allocation way that will always give the diffusion-reachable probe set item with the highest scores. Thus, these diffusion-reachable probe set items are ranked in the top of recommendation list. Finally, the theoretical upper bound of recommendation accuracy is obtained by measuring recommendation list. In practice, this perfect resource allocation method does not exist at present, but this Ideal method reveals the maximum recommendation accuracy that can be achieved by solely adjusting the resource allocation process in the diffusion-based recommendation methods. To characterize the predictability of this kind of algorithms, we employ ranking score\cite{9} (see definition in section \ref{my_section}) to measure the recommendation accuracy.

\begin{figure}[!htb]
	
	\centering
	\includegraphics[width=14.5cm]{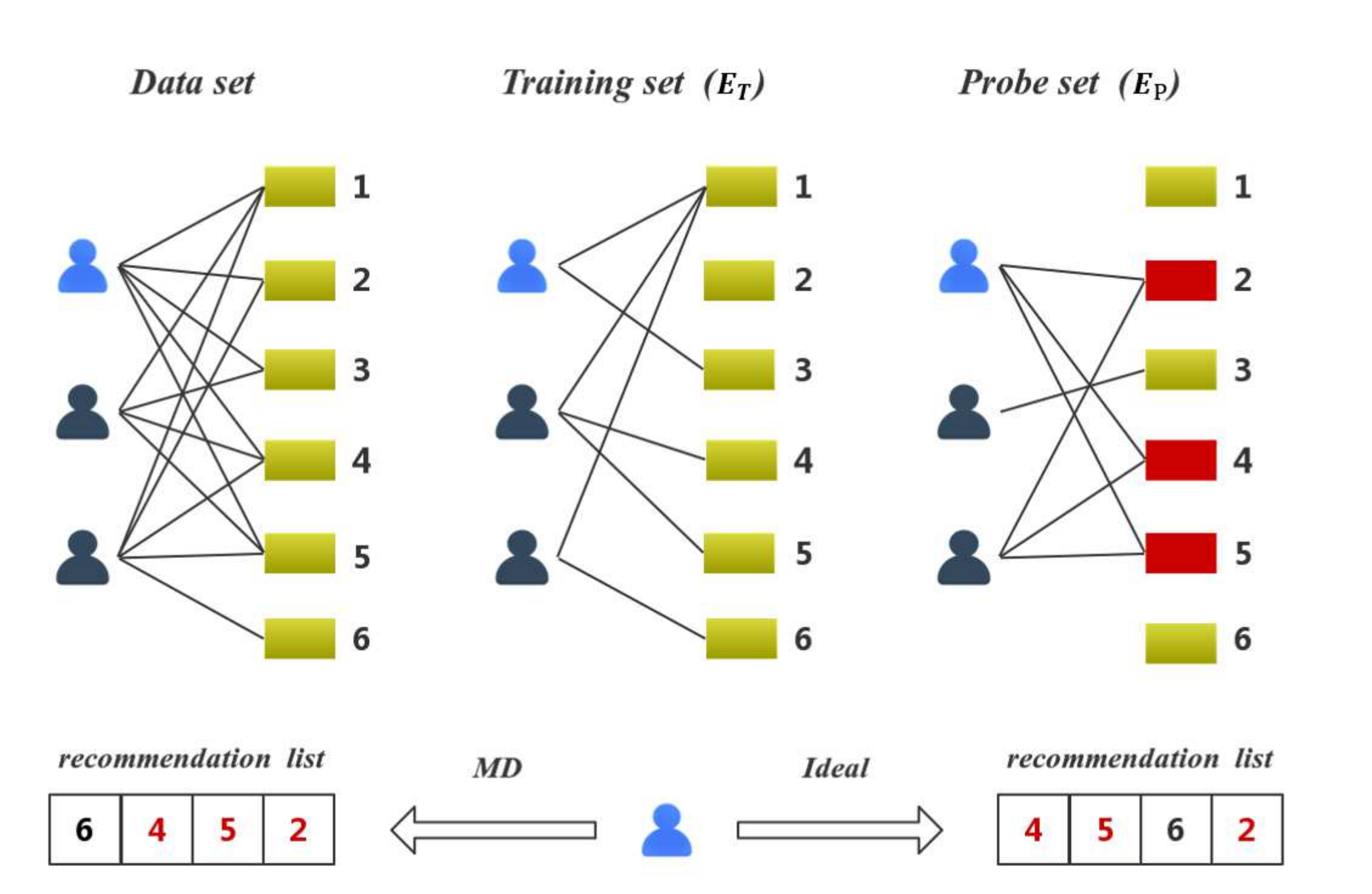}
	\caption{\textbf{A simple illustrative example of ideal method}. The user in wathet is the target user and rectangles denote items. The red rectangle denotes objects in probe set that will be bought by a target user. The recommendation list of the MD algorithm and the ideal method are shown, where the recommendation list of the ideal method is generated by top ranking the probe set objects whose resource can be reached by the 3 steps diffusion process. (For interpretation of the references to color in this figure legend, the reader is referred to the web version of this article.)}
	\label{Fig.1} 
	
\end{figure}

A simple example of this procedure is illustrated in \autoref{Fig.1}. The original data is divided into training set ($E_T$) and probe set ($E_P$). A training set is treated as a historical record about items purchased by users. A probe set is consider as a set of items that will be brought by users. In the dataset, the target user (wathet) has selected the $item_1$ and $item_3$ and will buy the $item_2$, $item_4$ and $item_5$. Here, we use the mass diffusion(MD)\cite{10} to recommend objects to target users (wathet). The result of recommendation for the target user is $[6,4,5,2]$ on the training set. At the same time, we use ideal method to obtain the recommendation list $[4,5,6,2]$. The ideal method is implemented by putting $item_4$ and $item_5$ in top of the recommendation list. The position of $item_2$ in the recommendation list is not changed because the resource cannot reach it by the three-step diffusion. We measure the prediction accuracy of recommendation algorithms by ranking score\cite{9}. The ranking score of MD and ideal is $0.75$ and $0.583$ respectively. Thus, the predictability of diffusion-based method is $0.583$.

\captionsetup[figure]{labelfont={bf},labelformat={default},labelsep=period,name={Fig.}}

\section{Experiment}
\label{section3}
We conduct several experiments to analyse the result of predictability. Seven datasets are chosen for the experiments: Movielens, Netflix, Amazon, Delicious, Douban, Epinions and Stack. We select the ranking score as the evaluation metric of recommendation accuracy.

\subsection{Datasets}

In the following simulation, we use two benchmark datasets and five subsets randomly selected from the original dataset: Movielens\cite{26}, Netflix\cite{27}, Epinions\cite{28}, Stack\cite{29} , Amazon\footnote{\url{https://www.amazon.com/}}, Delicious\footnote{\url{http://www.thedeliciousgroup.com/}} and Douban\footnote{\url{https://www.douban.com/}}. Movielens is a web-based personalized recommendation system to recommend movies for users. The Movielens data  consists of 100000 ratings (1-5) resulting from 943 users remarked 1682 movies and contains the seven-month period from September 19th 1997 to April 22nd 1998. Netflix is an American global provider of streaming films and television series. The Netflix data set is a subset of the original data set released for the purpose of the NetflixPrize. The data contains 1014 users and 1977 movies chosen from the original data at random and all links among them. The subset network contains 62848 links and time stamp information. Epinions is an online product rating site. Epinions dataset are made up of users and products. Each edge connects an user with a product and represents a rating. Stack Overflow is the main question and answer website of the Stack Exchange Network where nodes reprenst users and posts and an edge denotes that an user has marked a post as a favorite. Amazon, Delicious and Epinions dataset is available by crawing their website. The above six subsets are randomly selected from the original dataset. The detailed statistical characteristics of the seven networks are given in \autoref{Table1}. Obviously, the Netflix network is more sparse than Movielens network. Movielens and Netflix datasets are used to analyze the Figure 2 to 5. Fi are based on all datasets.

\captionsetup[table]{labelfont={bf},labelformat={default},labelsep=period,name={Table}}

\begin{table}[!htbp]
	\setlength{\abovecaptionskip}{3pt}
	\caption{The detailed statistical characteristics of the eight experimental dataset}
	\label{Table1}
	\centering
	\small
	\begin{tabular}{ccccccc}
		\toprule  %添加表格头部粗线
		Datasets& Users& Items& Links& $<k_{user}>$& $<k_{item}>$& Sparsity\\
		\midrule  %添加表格中横线
		Movielens& 943& 1,682& 100,000& 106.04& 59.45& $6.30\times10^{-2}$ \\
	
		Netflix& 1,014& 1,977& 62,848& 61.98& 31.79& $3.13\times10^{-2}$ \\
		
		Amazon& 8,071& 73,071& 101,155& 12.53&  1.38& $0.02\times10^{-2}$ \\
		
		Delicious& 1,000& 77,371&
		129,231& 129.23& 1.67& $0.17\times10^{-2}$\\
		
		Douban& 1,699& 59,102&
		195,372& 114.99& 3.31& $0.19\times10^{-2}$\\
		
		Epinions& 1,205& 106,685& 155,121& 128.73& 1.45& $0.12\times10^{-2}$\\
		
		Stack& 54,520& 38,904& 131,921&  2.42& 3.39&
		$0.06\times10^{-3}$\\
		
		\bottomrule %添加表格底部粗线
		
	\end{tabular}
\end{table}

\subsection{Datasets partition}
In order to analyze the result, we use two ways to divide datasets: random partition and temporal division. (1) Random datasets division refers to that all links in bipartite network are randomly divided into a training set($E_T$) and a probe set($E_P$) according to partition proportion. (2) Temporal dataset partition is based on time stamps on links (i.e. earlier links are put in the training set and later links are put in the probe set). The training set is viewed as known information, which is used to recommend objects to users. The probe set is used to verify the prediction accuracy. Obviously, $E_T \bigcap E_P = \o{}$ and $E_T\bigcup E_P =E$. All the simulation results are obtained by averaging over fifty independent experiments when the dataset is diveded randomly into training set and probe set.

\subsection{Evaluation metrics}
\label{my_section}

Various evaluation metrics are able to estimate the recommendation accuracy of diffusion-based methods. Here, we employ the ranking score to quantify the maximum recommendation accuracy of diffusion-based methods. In order to illustrate the result, we also propose a so-called ``diffusion coverage'' measure to quatify the range of resources diffusion.

(1) Ranking Score(RS)\cite{10}:
Ranking score measures the prediction accuracy of a recommendation algorithm which generates an ordered queen of all uncollected items for an arbitrary user to match the user's perference in the future. Ranking score evaluates the accuracy of predicition from the perspective of ranking. For a target user $u_i$, the recommendation list is generated by recommendation algorithm. Thus, for the edge $u_i-o_\alpha$ in probe set, we measure the position of object $o_\alpha$ in recommendation list of user $u_i$. A good recommendation algorithm is expected to give the object in probe set a higher rank, which leads to a small ranking score. Ranking score of a target user is obtained by averaging over all entries in probe set to quantify the recommendation accuracy of method. The specific formula is as follows:
\begin{equation}
RS_{u_i} =\frac{1}{|\{u_{i\alpha}\in E^P\}|}\sum_{u_{i\alpha} \in E^P}\frac{l_{i\alpha}}{L_{u_i}}
\end{equation}

where $RS_{u_i}$ denotes the ranking score of the user $i$. The $u_{i\alpha}$ is user $i$ - object $\alpha$ relations in the probe set and $l_{i\alpha}$ is the rank of object $\alpha$ in the recommendation list of user $i$. $L_{u_i}$ equals to $|O-E_{u_i}^T|$, namely the number of uncollected item for user $i$. $|\{u_{i\alpha}\in E^P\}|$ denotes the number of item for user $i$ in probe set. The $RS$ of the whole system is obtained by averaging $RS_{ui}$ over all users. Obviously, the range of $RS$ is that $RS \in (0,1)$. The smaller $RS$, the higher the prediction accuracy of recommendation algorithms, vice versa.

(2) Diffusion Coverage(DC):
Diffusion coverage quantifies the fraction of uncollected items that resource can be reached by the diffusion-based recommendation methods. DC is defined as the ratio of unselected objects covered by resources to the number of uncollected objects. For a target user, the greater the scope of resource diffusion, more products have the possibility to be recommended to users in this system and vice versa. The definition is as follows:
\begin{equation}
DC_{u_i} = \frac{n_{u_i}}{N_{u_i}}
\label{equation}
\end{equation}
where $n_{u_i}$ denotes the number of uncollected item that the resource can cover for user $i$. $N_{u_i}$ is the number of objects that have never been picked by the user $i$. Obviously, $DC_{u_i}\in(0,1)$. The larger the DC, the greater the resource diffusion coverage.

\section{Results}
\label{section4}

In this section, we compare the accuracy between diffusion-based algorithms and predictability, and analyze a serious of results. MD is employed to conduct analysis because MD is a classical diffusion-based method and recommendation accuracy of MD is very high. In the following results, note that the predictability of diffusion-based methods refers to the ranking score of Ideal method. 

\begin{figure}[htb]
	
	\centering
	\includegraphics[width=16.5cm]{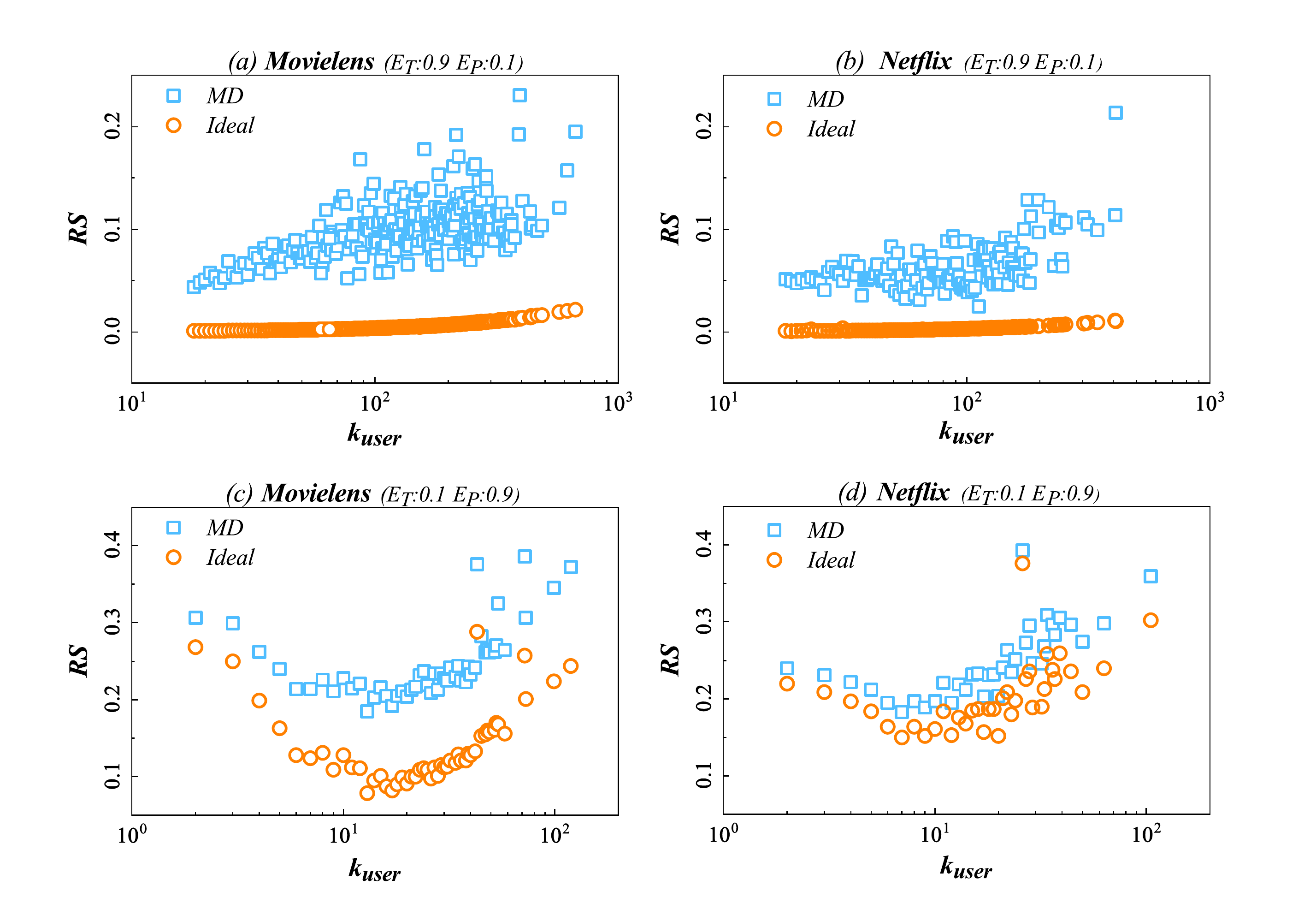}
	\caption{\textbf{The predictability of diffusion-based algorithms and the recommendation accuracy of mass diffusion}. The ranking score of both MD and Ideal method are shown for comparison. Points are averages over ranking score under the same degree of user. \textbf{a,b} The Movielens and Netflix datasets are divide into $E_T$ and $E_P$ respectively according to the ratio of 90\% to 10\%, this means we make recommendation on a dense network. \textbf{c,d} The Movielens and Netflix datasets are divide into $E_T$ and $E_P$ respectively according to the ratio of 10\% to 90\%, which denotes items is recommended for user on a sparse network.}
	\label{Fig.2}
	
\end{figure}

\subsection{The predictability of diffusion-based method}
The analysis of predictability on dense and spares network in studied Movielens and Netflix reveals three general results (see \autoref{Fig.2}): 
$(1)$ For users with different degree, the value of RS of Ideal method is close to 0 on a dense network.  This denote items to be bought by users can be accurately recommend when employing a diffusion-based algorithms with perfect way of resource allocation. In other words, the current recommendation accuracy of diffusion-based algorithms can be enhanced if the scoring scheme of the diffusion-reachable items is improved (see \autoref{Fig.2} a,b). 
$(2)$ The value of RS of Ideal method for both small and big degree user higher than other users on sparse network, where small / big degree user refer to user's degree less than 3 / greater than 54. This result suggest that the predictability of diffusion-based methods for these users is very low and it is difficult for small and big degree users to recommend items to be bought on a sparse network (see \autoref{Fig.2} c,d). 
$(3)$ For various users, the RS of MD is very close to RS of Ideal method on Netflix dataset (see \autoref{Fig.2} d). In this case, if we are dedicated to extend existing diffusion-based algorithms, the improvement of recommendation accuracy is not obvious. This results reveals that the improvement of recommendation accuracy based on diffusion algorithms on sparse network need to take other methods to increase its predictability instead of simply adjusting the resource allocation matrix.
Our study highlights the fact whether a new recommendation algorithms based on resource diffusion should be proposed need to depend on its predictability.

\begin{figure}[t]
	
	\centering
	\includegraphics[width=16.5cm]{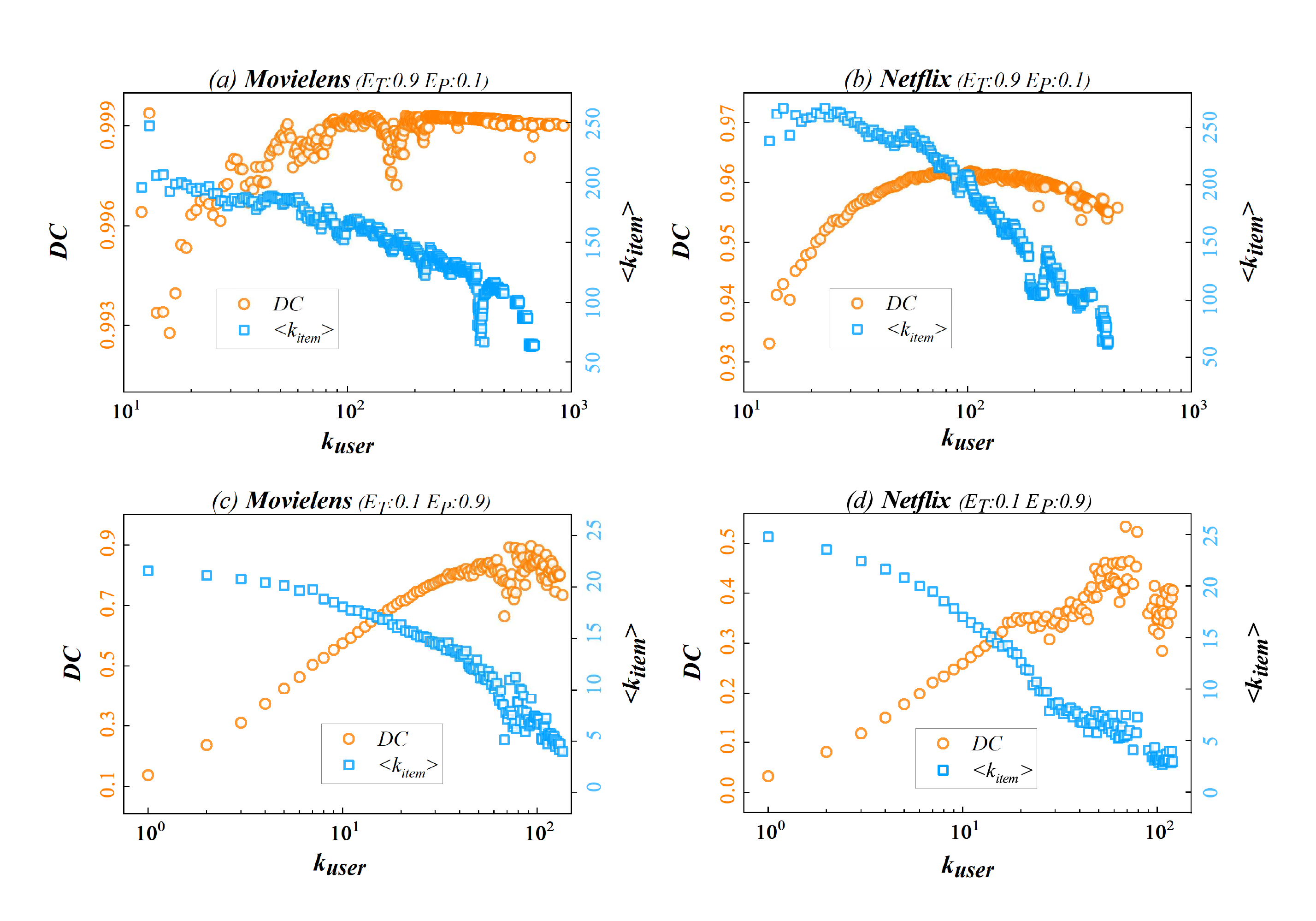}
	\caption{\textbf{The diffusion coverage and the recommended item average degree for users with different degree}. The data division setting as well as the averaging procedure are the same as those in \autoref{Fig.2}. \textbf{a,c} Movielens. \textbf{b,d} Netflix.}
	\label{Fig.3}
	
\end{figure}

Intuitively, the lower predictability of diffusion-based algorithms for small degree users is not surprising on a sparse network, because the link information of users is too less to spread the resource to unselected items. However, for big degree users, the predictability of diffusion-based algorithms is contradict to what we have imagined. Next, we propose two metrics to illustrate the mechanism behind it: diffusion coverage and item average degree (detailed information of diffusion coverage can be seen in \autoref{equation}). The item average degree reflects the popularity of item selected by users. \autoref{Fig.3} show the diffusion coverage and item average degree for various users on dense and sparse networks, which is conducted on Movielens and Netflix respectively. 
$(1)$ On a dense network (see \autoref{Fig.3} a), the smallest value of DC among all users reach 0.99, which suggesting resource almost fully cover items that have never been selected by users. In addition, the item average degree for all users is very large even though it decreases with user degree, revealing users tend to buy popular items that are covered easily by resource. This is a possible explanation for the higher predictability of diffusion-based method observed on dense network. 
$(2)$ On a sparse network (see \autoref{Fig.3} c), the lower DC for small degree users reflect the fact small degree users propagate initial resource to less unselected items. This could explain the lower predictability of diffusion-based methods. As for big degree users, the item average degree is very low although the DC is high. This is indicative of big degree users buy many cold items that seldom receive initial resource of target users. Although initial resource of big degree users cover a lot of items, the probability of obtaining resource for cold items is relatively low.  This leads to the predictability of diffusion-based recommendation algorithms is lower. Indeed, the predictability is best when the two curves crossover (see \autoref{Fig.3} c and \autoref{Fig.2} c). Similar interpretation on Netflix dataset is obtained form the \autoref{Fig.3} b,d. 
The proposition of two metrics help us understand deeply the limitations and advantage of diffusion-based recommendation algorithms.

\begin{figure}[!tb]
	
	\centering
	\includegraphics[width=16cm]{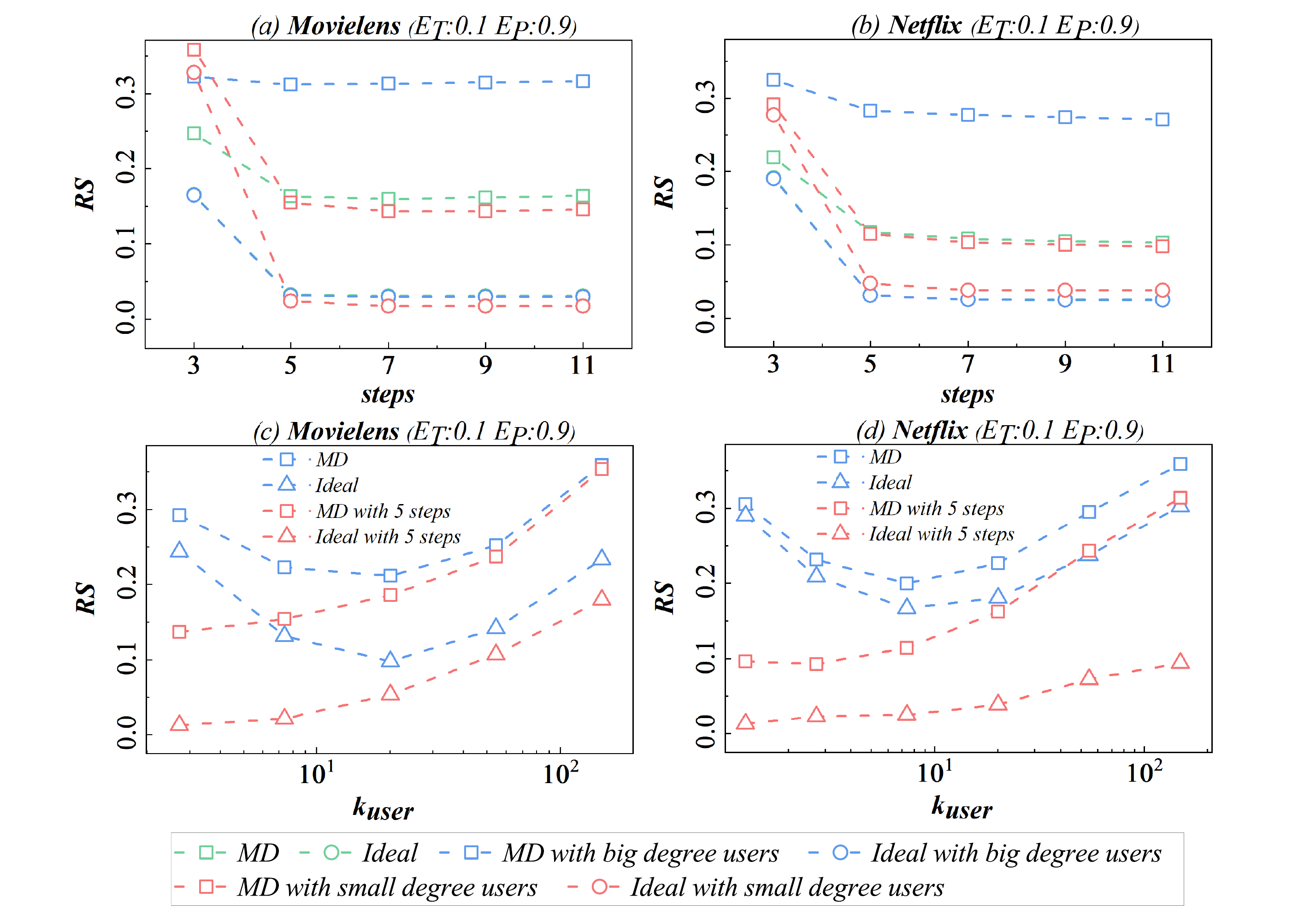}
	\caption{\textbf{The predictability of diffusion-based algorithms and ranking score of mass diffusion by muti-steps resource diffusion}. \textbf{a,b} The ranking score of mass diffusion and Ideal method with different diffusion steps on a sparse network. Here, small / big degree user refer to user's degree less than 3 / greater than 54 respectively. To clearly observe the improvement of accuracy under different conditions, we distinguish from user degree. Points are averages over ranking score under the same step. \textbf{c,d} Scatter plot of the ranking score as a function of user degree, which is conducted on Movielens and Netflix datasets respectively. The ranking score of mass diffusion and Ideal method with three steps and five steps are shown on a sparse network. Note that when diffusion step is three, the methods reduce to the classic recommendation methods in \cite{10}. Points are averages over ranking score binned in intervals of exponent.}
	\label{Fig.4}
		
\end{figure}

	\subsection{The improvement of predictability of diffusion-based methods} 
	The lower recommendation accuracy of MD observed from the \autoref{Fig.2} c,d, is mainly constrained by its predictability. We next turn our attention to improve the predictability of diffusion-based algorithms. Inspired by diffusion coverage, we enhance the predictability on sparse network by multi-steps resource diffusion. The \autoref{Fig.4} shows RS of MD and Ideal method with multi-steps diffusion, namely the accuracy of MD and predictability. The RS of different curves reach the lowest and keep stable when resource cover items with five step diffusion. We find the predictability can be improved significantly, especially for small degree users (see \autoref{Fig.4} a,b). This again support our explaination. In \autoref{Fig.4} c,d, we further compare the difference of accuracy between three steps and five steps. The result shows accuracy of MD for small degree users can be improved largely on sparse network. Actually, this is mainly because predictability of diffusion-based algorithms is enhanced by five steps diffusion.

\begin{figure}[!htb]
	
	\centering
	\includegraphics[width=16cm]{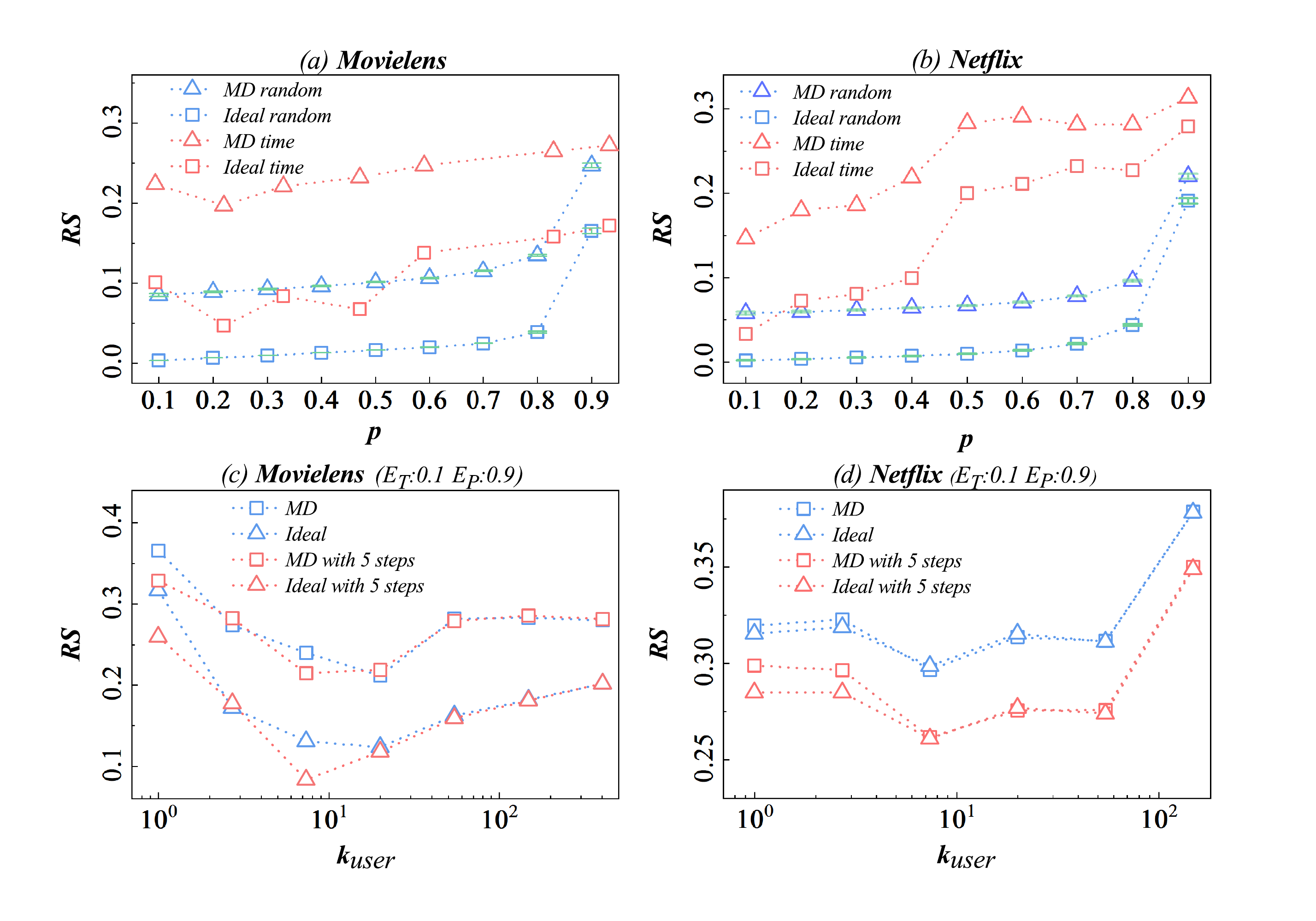}
	\caption{\textbf{The predictability of diffusion-based algorithms and ranking score of mass diffusion under temporal dataset partition}. \textbf{a,b} The ranking score with division ratio p, where p represents the ration of the probe set. Green bars represent the standard error. \textbf{c,d} Scatter plot of the ranking score as a function of user degree, which is conducted on Movielens and Netflix datasets respectively.  The ranking score of MD and Ideal method with three steps and five steps are also shown on a sparse network. The averaging procedure is the same as those in \autoref{Fig.4} \textbf{c,d}.} 
	\label{Fig.5}
	
\end{figure}

\subsection{The predictability of diffusion-based algorithms under temporal dataset partition} 
In the recommendation, it is more important to acquire the temporal information of items. For example, the taste of users changes over time. Thus, we analysis the predictability of diffusion-based methods according to the temporal dataset partition, and compare recommendation accuracy under different partition ways. \autoref{Fig.5} a,b show the RS of Ideal method and MD with various p under two partition ways, where p refers to the proportion of probe set. Obviously, the recommendation accuracy of MD and predictability under temporal partition is much lower than random division, suggesting the accuracy of recommendation in real online system is overestimated by random partition. Besides, the difference of accuracy between MD and predictability gradually decrease with partition proportion p, a sparser dataset Netflix in particular. This again reveals we only can seek to other ways to increase its predictability if we want to improve the recommendation accuracy of diffusion-based algorithms on a sparse network. To observe predictability for various users under temporal partition,  the RS of both MD and Ideal method with three and five steps diffusion is shown in \autoref{Fig.5} c,d, our finding is not specific to the result in \autoref{Fig.4} c,d, with the same conclusion. 

\begin{figure}[htb]
	
	\centering
	\includegraphics[width=16cm]{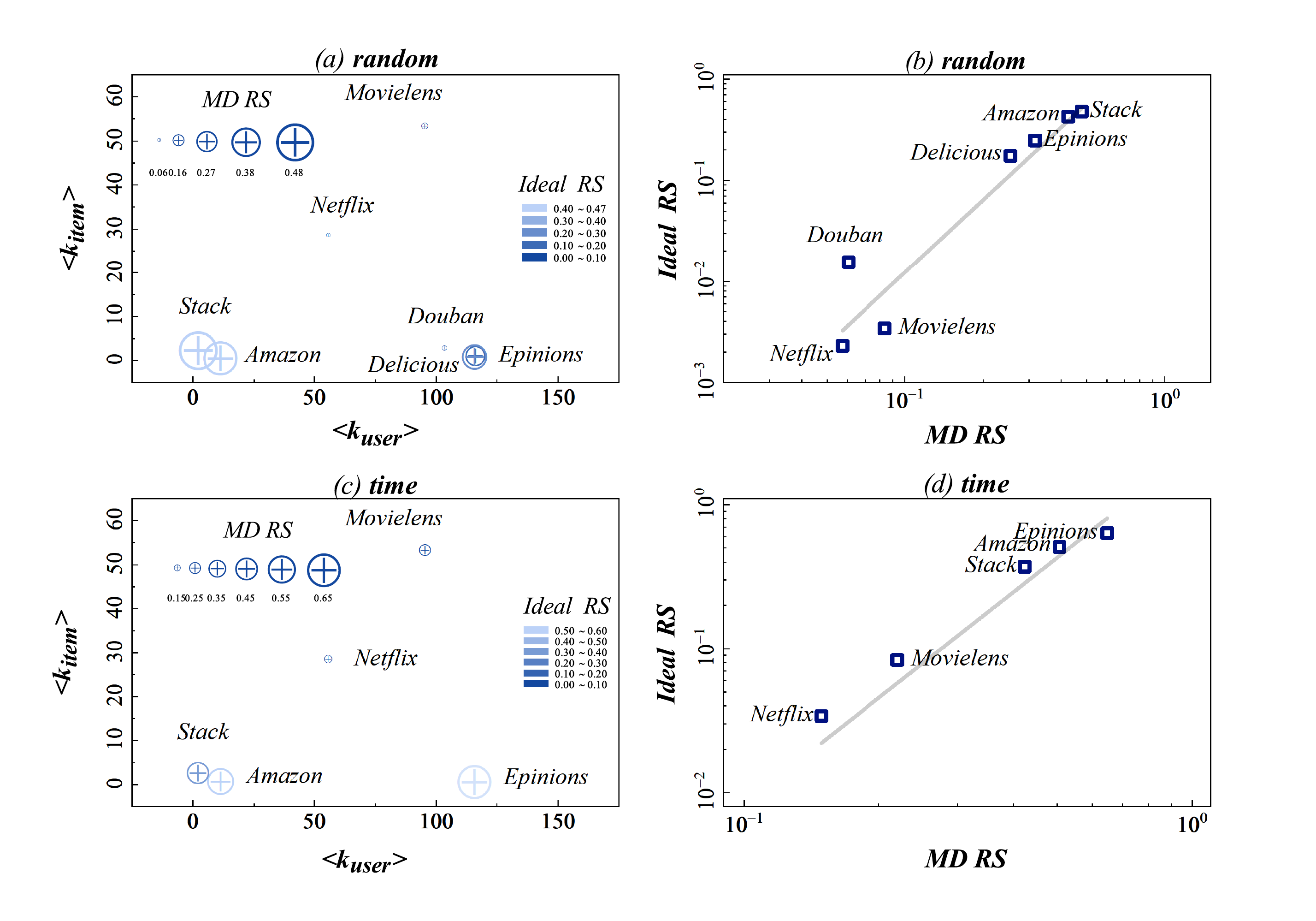}
	\caption{\textbf{The predictability of diffusion-based algorithms and ranking score of mass diffusion under different network datasets}. \textbf{a,c} The location of different dataset in the user average degree and item average degree space, with the size and color of each point as the ranking score of MD and Ideal methods in this dataset respectively. \textbf{b,d} The recommendation ranking score of Ideal diffusion method versus the ranking score of the MD method in different data set. \textbf{a,b} shows the results with random data division while \textbf{c,d} shows the results with temporal data division. A fitted linear line is plotted in each figure to guide eyes.}
	\label{Fig.6}
	
\end{figure}

In fact, the density of the link in network directly determine the prediction ability of diffusion-based recommendation algorithms. To better understand how network sparsity affects the predictability, two metrics is proposed characterize different network: average degree of user $\langle k_{user} \rangle$ and average degree of item $\langle k_{item} \rangle$. The average degree of user denotes that average number of items purchased by an user, suggesting the likelihood that an user will buy items. The average degree of item refers to the average number of purchase for one item, indicating the likelihood that an item is purchased by users. \autoref{Fig.6} a,c show RS of both Ideal method and MD for different kinds of dataset when datasets are embedded in the space of average degree of item and user. We find that the greater the average degree of users and items, the higher the predictability of diffusion-based algorithms. However, for those networks that have only higher average degree of users, the predictability of diffusion-based is uncertain. For instance, predictability in the Douban datasets is very high with random division, while the predictability in Epinions is relatively low. Moreover, the predictability under random partition higher than that of temporal division in all datasets, except for Stack. The point of Delicious is covered by Epinions (see \autoref{Fig.6} a, more detailed information can be see in \autoref{Table2}). Finally, we verify the effectiveness of Ideal method under different sparsity of datasets. The scatter plot of Ideal RS as function of MD RS in all datasets is show in \autoref{Fig.6} b,d. The result that RS of MD is positively correlated with RS of Ideal is not surprising, suggesting the recommendation accuracy of MD is determined by its predictability. But an absence of a correlation would suggest that the predictability of diffusion-based algorithms are internally nonsensible and would undermine the validity of Ideal method.

% \usepackage{multirow}
\begin{table}[!htbp]
	\setlength{\abovecaptionskip}{3pt}
	\caption{The ranking score of MD and Ideal diffusion method on the seven experiment datasets with random and temporal data partition}
	\label{Table2}
	\centering
	\small
	\begin{tabular}{clcccccccc}
		\hline
		\multicolumn{2}{c}{\multirow{2}{*}{Partition way}} & \multirow{2}{*}{Method} & \multicolumn{7}{c}{Datasets} \\ \cline{4-10}
		\multicolumn{2}{c}{} &  & Netflix & Movielens & Amazon & Epinions & Stack & Delicious & Douban \\ \hline
		\multicolumn{2}{c}{\multirow{2}{*}{Random}} & MD & 0.0573 & 0.0837 & 0.4261 & 0.3168 & 0.4813 & 0.2554 & 0.0605 \\
		\multicolumn{2}{c}{} & Ideal & 0.0023 & 0.0034 & 0.4243 & 0.2471 & 0.4739 & 0.1742 & 0.0155 \\
		\multicolumn{2}{c}{\multirow{2}{*}{Time}} & MD & 0.1487 & 0.2203 & 0.5069 & 0.6472 & 0.4238 & \rule[3pt]{0.8cm}{0.08em} & \rule[3pt]{0.8cm}{0.08em} \\
		\multicolumn{2}{c}{} & Ideal & 0.0039 & 0.0834 & 0.5058 & 0.6317 & 0.3701 & \rule[3pt]{0.8cm}{0.08em} & \rule[3pt]{0.8cm}{0.08em} \\ \hline
	\end{tabular}
\end{table}

\section{Conclusions}
\label{section5}

Although many studies have focused on improving the accuracy of diffusion-based algorithms, the extent to which network can be predicted by diffusion-based algorithms lack of understand. In this paper, we mainly study the maximum predictive ability of diffusion-based methods, namely predictability of diffusion-based algorithms. We propose an Ideal method to quantify this predictability, which allow us to evaluate existing diffusion-based algorithms and provide us a guideline for designing some new resource diffusion methods. Through comparing the accuracy between predictability of diffusion-based algorithms and mass diffusion, we find that the predictability is very high on a dense network and those items that users will buy can be accurately recommended if we continue to optimize resource allocation matrix. In contrast, on a sparse network, the recommendation accuracy of mass diffusion is very close to its predictability. In this case, it is ineffective to extend the existing diffusion-based algorithms to improve accuracy. Interestingly, the predictability for those big and small degree users preform very poor on a sparse network. Thus, we proposed diffusion coverage and item average degree to illustrate the mechanism behind it. The key insight of our finding is that those users that buy many items tend to purchase unpopular items that initial resource seldom cover, and the small degree users propagate the initial resource to less items. Inspired by this idea, we show that the predictability could be improved profoundly by multi-steps resource diffusion, especially for small degree users. Besides, we also find the predictability of diffusion-based method under temporal dataset partition is lower than that of random partition, revealing that recommendation accuracy in real online system is overestimated by random data partition. 

The predictability of diffusion-based method allow us to evaluate the recommendation algorithms based on resource diffusion on the same work. This is very instructive for proposing some new diffusion-based algorithms. For example, we could focus on enhancing the recommendation accuracy when accuracy of algorithms are far from predictability, because these algorithms have great potential to improve. However, for those algorithms whose accuracy are very close to the predictability, we only find other methods to improve the predictability. Here, we present a multi-steps diffusion method to improve the predictability of diffusion-based methods on a sparse network. Besides, there are still many other methods to achieve higher predictability, such as coupling social network of user or adding virtual edges. Moreover, predictability of diffusion-based algorithms can also extract the network skeleton structure\cite{36} and mining key links. We can remove those edges that cannot effect the predictability of recommendation algorithms so that the recommendation algorithms only need to process a small amount of data and the computational complexity will be reduced accordingly. Currently, we only explore the predictability of diffusion-based algorithms, while how network structure affect the predictability of diffusion-based method still unknown, and it is of great interest to both researchers and practitioners in the future. 

\section*{Author Contributions}
Conceptualization, A.Z; Formal Analysis, P.Z, L.X and A.Z; Software and Visualization and Writing-Original Draft Preparation, L.X; Writing-Editing and Supervision, P.Z and A.Z;
  	
\section*{Acknowledgements}
This work is supported by the National Natural Science Foundation of China (Grant No.61403037, No.61603046), the Natural Science Foundation of Beijing (Grant No.L160008).

\section*{Conflicts of Interest}
The authors declare no conflict of interest.

\bibliography{mybibfile}

\end{document}